\documentclass[11pt]{article}

\usepackage{fullpage}
\usepackage{mdframed}
\usepackage{mathpazo}

\pdfoutput=1

\usepackage{amsmath, amssymb, amsthm,verbatim, graphicx, xcolor,bbm}
\usepackage{stmaryrd} 
\usepackage{url}

\usepackage{mathrsfs}  

\usepackage{braket}

\makeatletter
\newtheorem*{rep@theorem}{\rep@title}
\newcommand{\newreptheorem}[2]{%
\newenvironment{rep#1}[1]{%
 \def\rep@title{#2 \ref{##1}}%
 \begin{rep@theorem}}%
 {\end{rep@theorem}}}
\makeatother

\newreptheorem{theo}{Theorem}

\def\1{\mathbbm{1}}

\usepackage{hyperref}

\begin{document}

\title{Information reconciliation for discretely-modulated continuous-variable quantum key distribution}

\author{Anthony Leverrier\thanks{Inria, anthony.leverrier@inria.fr}}

\date{\today}	
\maketitle

\begin{abstract}
The goal of this note is to explain the reconciliation problem for continuous-variable quantum key distribution protocols with a discrete modulation. Such modulation formats are attractive since they significantly simplify experimental implementations compared to protocols with a Gaussian modulation. Previous security proofs that relied crucially on the Gaussian distribution of the input states are rendered inapplicable, and new proofs based on the entropy accumulation theorem have emerged. Unfortunately, these proofs are not compatible with existing reconciliation procedures, and necessitate a reevaluation of the reconciliation problem. 
We argue that this problem is nontrivial and deserves further attention. In particular, assuming it can be solved with optimal efficiency leads to overly optimistic predictions for the performance of the key distribution protocol, in particular for long distances.
\end{abstract}

Any quantum key distribution (QKD) protocol involves a quantum procedure providing the honest parties access with two classical strings ${\bf a} = (a_1, \ldots, a_n) \in \mathcal{A}^n$ and ${\bf b} = (b_1, \ldots, b_n) \in \mathcal{B}^n$. Here $\mathcal{A}$ and $\mathcal{B}$ are arbitrary alphabets, which can be binary,  $\mathcal{A} = \mathcal{B} = \{0,1\}$, in the case of BB84 where the raw keys are simply bit strings, or real-valued $\mathcal{A} = \mathcal{B} = \mathbbm{R}$ for continuous-variable (CV) QKD protocols with a Gaussian modulation, or even hybrid $\mathcal{A} = \{1, -1\}$, $\mathcal{B} = \mathbbm{R}$ for some CV QKD protocols with a discrete modulation. 
To extract a secret key from these data, a crucial step is the \emph{information reconciliation} procedure during which the honest parties will exchange some public information on an authenticated public channel with the goal of obtaining a common bit string which will later be processed through privacy amplification. \\

The information reconciliation task shares similarities with the \emph{channel coding problem} where one tries to send reliable information over a noisy channel, but with some crucial differences that we discuss below. While this task is fairy straightforward for discrete-variable QKD protocols such as BB84, and has typically a limited impact on the secret key rate, the story is quite different for CV QKD. 
In fact, before 2002, it was widely believed that CV QKD could never distribute a secret key as soon as the transmittance of the channel was below $50\%$, corresponding to about 15 km of optical fiber. To overcome this limit, one should resort to \emph{reverse reconciliation}~\cite{GG02b}, which means that the common bit string on which Alice and Bob agree should be a function ${\bf x}$ of ${\bf b}$ and that Bob should send some information ${\bf c}$ to Alice on a public channel, allowing her to recover ${\bf x}$ from the knowledge of ${\bf a}$ and ${\bf c}$. 
This problem turned out to be more challenging than anticipated for CV QKD. In fact, the difficulty to solve this problem optimally severely limited the range of experimental demonstrations of CV QKD for a long time, reaching about 25 km in 2007~\cite{LBG07}, and it was even argued that the complexity of the problem would make CV QKD unviable for long distances~\cite{ZGC08}. The main challenge is that the data shared by the honest parties becomes extremely noisy when the distance is large, making it hard to extract a common string optimally\footnote{In the case of BB84, most of the photons are not received by Bob, leading to some form of postselection. Only the detected photons are taken into account for ${\bf a}$ and ${\bf b}$ and the two bit strings are typically highly correlated with an error rate of only a few percent.}.
This problem can be bypassed by moving from a Gaussian modulation, where each $a_i$ is chosen independently from a Gaussian distribution, to a discrete modulation, where $a_i$ is chosen from $\{1, -1\}$ for instance. In this setting, one can achieve reconciliation almost optimally even in the high-noise regime relevant to long distances~\cite{LG09}. With this approach, distributing secrets key over more than 100 km with CV QKD becomes possible, at least if one neglects finite-size effects. Unfortunately, this only applies to some specific CV QKD protocols, for which there is no full security proof available at the moment. More precisely, we only know how to compute the asymptotic key rate valid against collective attacks~\cite{DBL21}.\\

Very recently, the entropy accumulation technique~\cite{DFR20} was applied to specific CV QKD protocols with a discrete modulation, establishing their security against general attacks~\cite{KGL23,BGW23}. In these protocols, Bob immediately discretizes the outcome of his coherent detection to get a single bit (per quadrature), and these bits will form the raw key. The catch is that the precise value of his measurement results cannot be exploited for information reconciliation, requiring to implement potentially less efficient reconciliation schemes than in~\cite{LG09}. 
We refer to the two kinds of information reconciliation schemes as \emph{hard-decision} (HD) for the ones where Bob only has access to a discrete variable (typically a single bit corresponding to the sign of the measurement outcome), and \emph{soft-decision} (SD) for those where Bob does not discretize his measurement outcomes and can exploit their absolue value for reconciliation.
We will argue that achieving high-efficiency reconciliation in the high-noise regime relevant to long-distance QKD is much easier for SD reconciliation. In particular, one should be careful when discussing their respective performance and should avoid making overly optimistic prediction by applying results from the SD version to the HD one. \\

The outline of this manuscript is as follows. In Section \ref{sec:coding}, I review the channel coding problem of transmitting information reliably on a noisy classical channel. Section \ref{sec:red} explains how the reconciliation problem can be reduced to a channel coding problem. Section \ref{sec:efficiency} discusses the performance of reconciliation schemes in the relevant high-noise regime.

\section{The channel coding problem}
\label{sec:coding}

The main strategy behind information reconciliation is to reduce the problem to a channel coding task over a specific class of (classical) channels, and to rely on the error correcting codes and decoders developed for these channels. Two families of channels are particularly relevant for CV QKD with a discrete modulation: the binary symmetric channel (BSC) and the binary-input additive white Gaussian-noise (BI-AWGN) channel,
\begin{itemize}
\item binary symmetric channel with crossover probability $p \in [0, 1]$:
\begin{itemize}
\item input: $X \in \{0, 1\}$,
\item output: $Y\in \{0, 1\}$ with $\mathrm{Pr}[Y=X] = 1-p$, $\mathrm{Pr}[Y\ne X]=p$.
\end{itemize}

\item BI-AWGN channel with signal-to-noise ratio (SNR) $s> 0$:
\begin{itemize}
\item input: $X \in \{-1, 1\}$,
\item output: $Y = X+Z$ with $Z \sim \mathcal{N}(0, 1/s)$. 
\end{itemize}

\end{itemize}
Here $\mathcal{N}(0, 1/s)$ is a centered Gaussian random variable with variance $1/s$. \\

In the channel coding problem, the goal is to transmit as much information as possible in a reliable fashion with $n$ uses of the channel. The general method is to pick an error correcting code encoding $k$ logical bits within $n$ physical bits, and to send a message corresponding to one of the $2^k$ possible codewords. The maximal theoretical value of the \emph{rate} $R = k/n$ for which reliable communication is possible corresponds to the \emph{channel capacity}, and is known as the \emph{Shannon limit}. The capacities $C_\text{BSC}(p)$ and $C_{\text{BI-AWGN}}(s)$ of the two channel families above are well known:
\begin{align}\label{eqn:capa}
C_\text{BSC}(p) = 1-h(p), \qquad C_{\text{BI-AWGN}}(s) = - \int \phi_s(x) \log_2 \phi_s(x) \, dx + \frac{1}{2} \log_2\left( \frac{s}{2\pi e}\right)
\end{align}
with 
\begin{align}
h(p) := -p\log_2 p - (1-p) \log_2(1-p), \qquad \phi_s(x) = \sqrt{\frac{s}{8\pi}} \left( e^{-s(x+1)^2/2} + e^{-s(x-1)^2/2}\right).
\end{align}
For small values of $s$, the capacity of the BI-AWGN channel is well approximated by that of the AWGN channel: 
\begin{align}\label{eqn:awgn}
C_{\text{AWGN}}(s) = \frac{1}{2} \log_2(1 + s).
\end{align}
What Shannon proved is that for $n$ sufficiently large, and for a rate $k/n$ below the capacity of the channel, one can reliably send $k$ bits of information with $n$ uses of the channel. 
This statement is only valid asymptotically, and for a finite value of $n$ with a specific error correcting code and decoding algorithm, the rate $k/n$ needs to be strictly smaller than the capacity. In addition, the proof of Shannon's theorem involves random codes which cannot be decoded efficiently. In the QKD scenario, one requires instead a very fast reconciliation procedure, able to process gigabits of data per second. This will in general lead to even smaller values of the rate used in practical implementations.

\section{Reducing the information reconciliation task to the channel coding problem}
\label{sec:red}

Let us focus on the case of a quadrature phase-shift keying (QPSK) modulation as in \cite{LG09}: Alice prepares coherent states of the form $|\alpha (a_1 + i a_2) \rangle$ and sends them to Bob who performs heterodyne detection, thus obtaining some complex number $b_1 + i b_2$. Here, we have $a_1, a_2 \in \{-1, 1\}$ and we are interested in the case of a Gaussian channel giving 
\[ b_k = a_k + z_k,\]
where $z_k \sim \mathcal{N}(0, 1/s)$. 
In practice, this approximates well experimental data since the channel between Alice and Bob is an essentially noiseless optical fiber, and that the main source of noise is the shot noise, which is accurately modeled by a Gaussian random variable. In particular, for a CV QKD experiment where the transmittance of the channel between Alice and Bob is $T$, the SNR is given by $s \approx \frac{T \alpha^2}{2}$, with a factor $1/2$ due to the balanced beamsplitter needed to measure both quadratures. For instance, at 100 km, we obtain $T=10^{-2}$ and typical values of $\alpha$ around $0.4$~\cite{LUL19,GGD19}, which give $s < 10^{-3}$. This is therefore the kind of SNR we need to consider for distributing secret keys over 100 km. \\

As mentioned above, in the case of reverse reconciliation, the goal is to find a function of $b_k$ that can be recovered by Alice by exploiting some additional side-information sent by Bob on a perfect authenticated classical channel. 
The natural function that is considered in most cases is simply the sign of $b_k$. There are two main strategies considered in the literature to reduce the reconciliation procedure to a channel coding problem: in the hard-decision (HD) strategy, Bob ignores the absolute value of $b_k$, while this value is carefully exploited in the soft-decision (SD) strategy. 
Given some $(a,b)$ as above, the reduction proceeds by defining a couple $(X, Y)$ that corresponds to the input/output of a particular channel. 

\begin{itemize}
\item \textbf{Hard-decision (HD) strategy} (e.g.~\cite{LUL19}):
\begin{itemize}
\item input $X = 0$ if $b\geq 0$ and $X=1$ if $b<0$,
\item output $Y = 0$ if $a=1$ and $Y=1$ if $a=-1$.
\end{itemize}
\item \textbf{Soft-decision (SD) strategy} (e.g.~\cite{LG09}):
\begin{itemize}
\item input $X = \frac{b}{|b|} \in \{-1, 1\}$,
\item Bob sends the side-information $|b|$ to Alice,
\item output $Y = |b| a \in \mathbbm{R}$.
\end{itemize}
\end{itemize}

Interestingly, the two resulting classical channels $X \to Y$ correspond respectively to a BSC and a BI-AWGN channel. \\

For the HD strategy, one recovers a BSC with crossover probability $p(s)$ given by
\begin{align}
p(s) = \sqrt{\frac{s}{2\pi}} \int_1^\infty e^{-sx^2/2} dx = \frac{1}{\sqrt{\pi}} \int_{\sqrt{\frac{s}{2}}}^\infty e^{-x^2} dx = \frac{1}{2} \left(1- \mathrm{erf}\left( \sqrt{\frac{s}{2}}\right) \right),
\end{align}
where $\mathrm{erf}$ is the error function. In particular, for small values of $s \ll 1$, corresponding to long distances, this probability can be approximated by
\begin{align}\label{eqn:approx}
p(s) \approx \frac{1}{2} - \sqrt{\frac{s}{2\pi}}.
\end{align}
For instance, at 100 km and $s=10^{-3}$, we obtain $p(0.001) \approx 0.487$.\\

On the other hand, the SD strategy is similar to a BI-AWGN channel. To see this, let us consider the variable $Z = Y-X$:
\begin{align*}
Z = |b| a - \frac{b}{|b|}
= \frac{b}{|b|} \left( ab -1\right)
=  \frac{b}{|b|} \left( a (a+z) - 1\right)=  \frac{ab}{|b|}  z
\end{align*}
where we used that $a^2 =1$ in the last equality. The absolute values of $Z$ and $z$ follow the same distribution, and the sign of $Z$ is $+1$ or $-1$ with probability $1/2$, independently on the value of $|Z|$. This shows that $Z$ is a Gaussian variable of variance $1/s$, independent of $X$. In other words, $X$ and $Y$ can be interpreted as the input and output of a BI-AWGN channel of SNR $s$.\\

With both strategies, we can therefore map the reconciliation task to a problem close to channel coding, either for the BSC or for the BI-AWGN channel. There is one catch, however. Here the input $X$ corresponds to the sign of $b$ and is therefore uniformly distributed. Recall that in the channel coding problem, the sender chooses the input as they wish. 
To solve this issue, the idea is that Bob will compute the \emph{syndrome} of ${\bf X} = (X_1, \ldots, X_n)$ for the error correcting code Alice and Bob agreed on, and will send this syndrome to Alice. Then Alice can simply perform the decoding in the code coset with the corresponding syndrome~\cite{Wyn75}. The number of bits sent during the reconciliation procedure, typically called the \emph{leakage} in the literature is $n-k = n(1-k/n)$:
\begin{align}\label{eqn:leak}
\mathrm{leak} = n \left(1-C + \left(C-\frac{k}{n}\right)\right),
\end{align}
where $C$ is the capacity of the relevant channel (binary symmetric or BI-AWGN).  The first term $n(1-C)$ is the minimum possible value of the leakage, for a scheme reaching the Shannon bound. The extra term $n(C - k/n)$ is due to the imperfect reconciliation, and may cause the QKD protocol to perform much worse than an ideal scheme with perfect reconciliation efficiency.
Now that we have mapped the reconciliation procedure to a channel coding problem, the question is how much additional leakage is due to the impossibility of reaching exactly the Shannon bound, and how it affect the secret key rate of the QKD protocol.

\section{Efficiency of the reconciliation procedure in the high-noise regime}
\label{sec:efficiency}

We focus here on the asymptotic secret key rate formula for collective attacks, which means in particular that we ignore all finite-size effects. With these assumptions, we can express the asymptotic secret key rate $K$ of the protocol (corresponding to the final key length divided by the number of channel uses) as
\begin{align}
K = K^{\text{ideal}} - \Delta_{\text{rec}}
\end{align}
where $K^{\text{ideal}}$ corresponds to the secret key rate when assuming that the reconciliation procedure reaches the Shannon bound, and the term 
\begin{align}
\Delta_{\text{rec}} = C-R
\end{align}
quantifies the imperfection of the reconciliation. Here $R = k/n$ is the rate of the classical code appearing in the reconciliation procedure. 
In general, the channel capacity $C$ is a function of the SNR $s$ and $R$ is chosen below the capacity so that the reconciliation procedure succeeds with reasonable probability. 
This means that $\Delta_{\text{rec}}$ is also a function of $s$, and our goal is to understand its behaviour, notably in the low-SNR regime $s \ll 1$ if we consider key distribution over long distances.\\

A standard figure of merit in the context of CV QKD is the \emph{reconciliation efficiency} $\beta$ defined as
\begin{align}\label{eqn:beta}
\beta = \frac{R}{C}.
\end{align}
The correction term to the ideal key rate then reads $\Delta_{\text{rec}} = (1-\beta)C$ and values of $\beta$ around $90$ to $95\%$ are typically quoted in the literature.\\

Given its importance for the performance of CV QKD, there has been a lot of work devoted to improving the speed and efficiency of information reconciliation of continuous variables: see \cite{YYY23} for a recent review. 
In particular, two main families of schemes were developed for CV QKD protocols \emph{with a Gaussian modulation}: slice reconciliation~\cite{VCC04} and the multidimensional reconciliation~\cite{LAB08}. The former amounts to discretizing Bob's data and is similar to the HD reconciliation schemes mentioned in the previous section; multidimensional reconciliation is similar to the SD schemes. In particular, all the literature discussing slice reconciliation for CV QKD is restricted to channels with reasonably large capacity, i.e.~SNR greater than 1. For such channels, reconciliation efficiencies can indeed reach $95\%$. Similar or better efficiencies are also achievable at much lower SNR, including the ones required for 100 km, for the multidimensional reconciliation.\\

When moving to CV QKD protocols with a discrete modulation, the difference between both approaches becomes much sharper, as so far in the literature, only the case of the SD strategy has been addressed in detail. 
We first discuss this case, which amounts to channel coding on the BI-AWGN channel, before addressing the case of the HD strategy.

\subsection{SD strategy: the BI-AWGN channel}

First, it should be noted that the BI-AWGN channel is one of the most relevant ones in classical communications and unsurprisingly much effort has been dedicated to designing good codes and decoders for this channel. 
However, the case of very noisy channels was not addressed much because it is more practical (and possible!) to work as sufficiently high SNR. A rare exception is CV QKD where one cannot amplify the signal since this would obviously add too much noise to allow for the extraction of a secret key. 
In particular, most codes discussed in the literature have rates typically exceeding $0.1$, which are much higher than those we wish to consider. 
Some works have considered codes for lower rates, for instance~\cite{AT12}, \cite{LLW03}, \cite{LGL06}. In particular, Ref.~\cite{LGL06} designs codes at rates around $0.01$ reaching efficiencies $\beta \approx 90\%$. \\

Crucially, from a good code for the BI-AWGN channel at low rate $R\leq 0.1$, it is easy to design good codes at rate of the form $R/m$ for any integer $m$ using the idea of \emph{repetition coding}. This idea was exploited for instance in~\cite{LG09} as the main ingredient allowing CV QKD to reach long distances. 
The idea is very simple: assuming the existence of a code with rate $R$ and good performance over the BI-AWGN channel of SNR $s$, i.e.~achieving some reconciliation efficiency $\beta(s) = R/C(s)$, one designs a new code that can be used for a BI-AWGN channel of SNR $s/m$. The new code is $m$ times longer with each bit of the original code is repeated $m$ times. To decode, one simply adds $m$ successive channel outputs to form a new variable. This variable corresponds to the output of a BI-AWGN channel with a variance $1/(m (s/m) )=1/s$. In other words, this simple trick has transformed the initial channel with SNR $s/m$ into one with a SNR $m$ times larger. 
Now, we can use the fact that the capacity of the BI-AWGN channel is essentially linear in $s$ at low SNR to conclude that this repetition code scheme does not lose anything in terms of capacity:
\begin{align}
C_{\text{BI-AWGN}}(s/m)  \approx  C_{\text{AWGN}}(s/m) = \frac{1}{2} \log_2\left(1+\frac{s}{m}\right) \approx \frac{1}{2m} \log_2\left(1+s\right) \approx \frac{1}{m} C_{\text{BI-AWGN}}(s).
\end{align}
In particular, we obtain a reconciliation efficiency at SNR $s/m$ given by
\begin{align}
\beta(s/m) = \frac{R}{m} \frac{1}{C(s/m)} \approx \frac{R}{C} = \beta(s).
\end{align}
Here the rate is replaced by $R/m$ since we need $m$ uses of the channel to transmit a bit. 
As an example, the multi-edge LDPC code of \cite{RU02} with rate $1/10$ decodes reasonably well the BI-AWGN channel of SNR $s=0.17$, which gives $\beta(0.17) = \frac{R}{C} \approx 0.88$. Then with the approach described above, we get $\beta(0.17/m) \geq 0.8$ for any value of $k$. Starting from a better code, for instance one from~\cite{LGL06} would lead to better reconciliation efficiencies. \\

Another advantage of this approach for coding over channels with very low SNR is that the decoding complexity is excellent since adding $m$ symbols together is as simple as it gets. In particular, only the initial code designed for SNR $s$ needs to be decoded, with a length independent of the number of channel uses in the limit of long distance.\\

This approach is applicable to the soft-decision strategy for reconciliation and works for the QPSK modulation scheme. It is not known however how to extend it to arbitrary modulation formats such as quadrature amplitude modulation (QAM). For modulations with a large constellation that approximate reasonably well a Gaussian modulation, for instance 64-QAM and 256-QAM as considered in~\cite{DBL21},  one should instead rely on the techniques developed for the Gaussian modulation. These can also achieve very high efficiencies but with more complicated strategies than repetition coding: see \cite{YYY23} for references. 
It is also possible to apply the soft-decision strategy to intermediate modulation schemes that rely on constellations defined over 1, 2 or 4 optical modes, as discussed in~\cite{LG11}. For such modulations, Alice sends for instance four successive coherent states $|\alpha a_1\rangle |\alpha a_2\rangle|\alpha a_3\rangle|\alpha a_4\rangle$ with the 4-dimensional complex vector $(a_1, a_2, a_3, a_4)$ chosen from a spherical code~\cite{EZ01,eczoo_spherical}, that is a finite set of vectors of unit length on the sphere of of $\mathbbm{C}^4$. In this case, one can exploit the multidimensional reconciliation approach~\cite{LAB08} to map the problem to a channel coding problem with inputs chosen from this constellation and a Gaussian channel. Such modulation schemes are expected to considerably improve the tolerance to excess noise compared to the QPSK modulation.

\subsection{Hard-decision strategy: the BSC}

To establish the security of a CV QKD protocol, it is convenient that Bob discretizes immediately its outcome. One can then rely on powerful approaches such as the entropy accumulation theorem to get a lower bound on the conditional smooth min-entropy of the raw key (corresponding to Bob's discrete outcome), and thus obtain full security proofs. Generalizing this to the case where Bob's outcome is continuous (or discretized with a very large alphabet) is more complicated and remains open at the moment. 
This is one of the reasons why recent works such as~\cite{KGL23,BGW23} consider protocols where Bob discretizes his data in a small number of bins. \\

As discussed in the previous section, the reconciliation problem reduces in this case to a channel coding task for the BSC, but with a very large crossover probability. Recall that at 100 km, the typical crossover probability is greater than $0.48$. Unfortunately, the question of designing good codes and decoders for such channels has not been investigated much in the literature. In particular, codes with high performance over the BSC were studied in the context of discrete-variable QKD, but only for small crossover probabilities of at most 10 to 20 $\%$. This is because larger values do not allow the honest parties to extract a secret key.\\

\begin{figure}[h!]
\begin{center}
\includegraphics[width=0.7\linewidth]{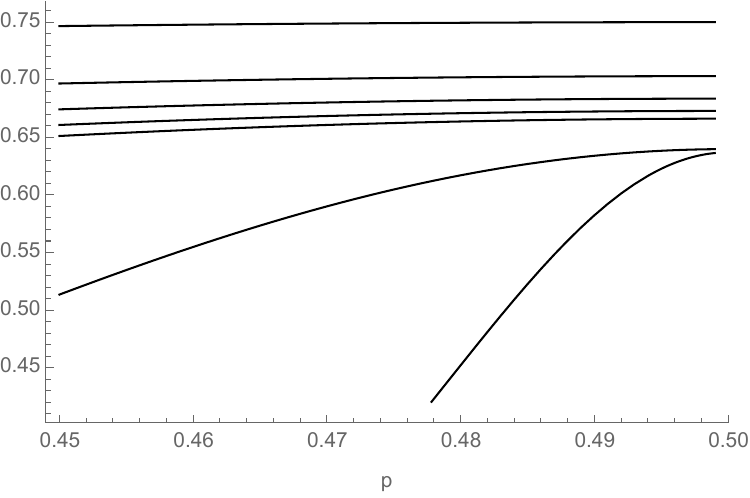}
\end{center}
\caption{Loss of reconciliation efficiency $\frac{\beta(p)}{\beta(p_m)}$ computed with \eqref{eqn:ratio} for repetition coding with the BSC of parameter $p$, with $m$ repetitions. From top to bottom: $m = 3,5,7,9,11, 101, 1001$.}
\label{fig}
\end{figure}

One could ask whether the idea of repetition coding that works so well for the BI-AWGN channel is also applicable here. 
In fact, one can use it, but at the cost of an important loss in terms of channel capacity. 
Recall that the capacity of the BSC of parameter $p$ is $C(s) = 1 - h(p)$. If we combine $m$ outputs of the BSC of parameter $p$ to get a single, less noisy bit, we are obtaining a new BSC with crossover probability
\begin{align}
p_m = \sum_{\ell=0}^{\left\lfloor m/2 \right\rfloor} \tbinom{m}{\ell} p^{m-\ell} (1-p)^{\ell}.
\end{align}
The reconciliation efficiency $\beta(p)$ obtained with this scheme is related to the reconciliation efficiency $\beta(p_m)$ corresponding to the better channel through 
\begin{align}\label{eqn:ratio}
\frac{\beta(p)}{\beta(p_m)}  = \frac{1}{m} \frac{1-h(p_m)}{1-h(p)}.
\end{align}
This ratio is displayed on Fig.~\ref{fig}, and is always significantly below 1. This shows that repetition coding will not help here. For instance, if the physical crossover probability is $p=0.48$, then repeating about 1000 times gives $p_{1000} \approx 0.10$. This means that one could use a code optimized for the BSC of parameter $0.1$, maybe with a reconciliation efficiency of $95\%$, but at the cost of an extra factor of $ \frac{1}{m} \frac{1-h(p_m)}{1-h(p)}$, which is about $0.45$ for $m=1000$. This gives a reconciliation efficiency well below 50$\%$, incompatible with key extraction at long distance. 
If a good code is available for the BSC with crossover probability $0.2$, then it is sufficient to take $m \approx 440$ and the loss in reconciliation efficiency due to repetition coding becomes around $0.55$, still too low to be helpful.\\

While these facts do not forbid the existence of reconciliation schemes with high efficiency for the BSC with large crossover probablity, they show that one cannot rely on very simple techniques such as repetition coding to obtain large families of codes in the high-noise regime. Rather one will likely need to design new codes for every value of the crossover probability.
One possibility is to use polar codes, which are known to be capacity achieving. This means that for a sufficiently large block size, they can in principle achieve any desired reconciliation efficiency. Such codes have been considered for the reconciliation of BB84 in~\cite{JK14} for instance, but already require length of $2^{24}$ to achieve $\beta \approx 0.95$ for a low crossover probability of $11 \%$.

 \section{Conclusion}
 \label{sec:open}
 
There are currently two main approaches for CV QKD with a discrete modulation, depending on whether Alice and Bob only exploit the sign of Bob's measurement outcomes or also their absolute value. In the first case, it is possible to prove the security of the protocols against the most general attacks, while this remains open at the moment for protocols where Bob exploit the absolute value of his measurement outcomes. 
The price to pay for this better security status is that the reconciliation problem is significantly more difficult, and that current coding strategies do not appear to yield reconciliation efficiencies sufficient to distribute secret key over long distances. Whether or not it possible to get the best of both worlds -- protocols reaching long distances with complete security proofs -- is certainly one of the most intriguing theory questions for CV QKD.


\end{document}